\documentclass[floatfix,twocolumn,aps,prl,showpacs,amsmath,amssymb]{revtex4-1}

\usepackage[latin1]{inputenc}
\usepackage[dvips]{graphicx}
\usepackage{dcolumn}
\usepackage{bm}
\usepackage{dsfont}
\usepackage{color}
\usepackage{url}
\usepackage[colorlinks=true ,citecolor=blue]{hyperref}
\usepackage{amsmath,amssymb}

\definecolor{Nathanblue}{rgb}{0.,0.24,0.51}

\def\bs#1{\boldsymbol{#1}}

\newcommand{\be}{\begin{equation}}
	\newcommand{\ee}{\end{equation}}
\newcommand{\bq}{\begin{eqnarray}}
	\newcommand{\eq}{\end{eqnarray}}

\begin{document}

\title{Revealing tensor monopoles through quantum-metric measurements}

\author{ Giandomenico Palumbo and Nathan Goldman}
\affiliation{Center for Nonlinear Phenomena and Complex Systems,
	Universit\'e Libre de Bruxelles, CP 231, Campus Plaine, B-1050 Brussels, Belgium}

\date{\today}

\begin{abstract}

Monopoles are intriguing topological objects, which play a central role in gauge theories and topological states of matter. While conventional monopoles are found in odd-dimensional flat spaces, such as the Dirac monopole in three dimensions and the non-Abelian Yang monopole in five dimensions, more exotic objects were predicted to exist in even dimensions. This is the case of ``tensor monopoles", which are associated with generalized (tensor) gauge fields, and which can be defined in four dimensional flat spaces. In this work, we investigate the possibility of creating and measuring such a tensor monopole, by introducing a realistic three-band model defined over a four-dimensional parameter space. Our probing method is based on the observation that the topological charge of this tensor monopole, which we relate to a generalized Berry curvature, can be directly extracted from the quantum metric.
We propose a realistic three-level atomic system, where tensor monopoles could be generated and revealed through quantum-metric measurements. 

\end{abstract}

\maketitle

\emph{Introduction: } Magnetic monopoles were originally introduced by Dirac in 1931~\cite{Dirac}, in view of  proving the quantization of the electric charge in quantum electrodynamics. While the Dirac monopole has played an important role in high-energy physics, in particular due to its topological nature~\cite{WuYang_PRD,Wu-Yang}, an experimental confirmation of its existence is still lacking. Since Dirac's original work, a zoo of monopoles have been identified in the context of gauge theory. Prominent examples are the 't Hooft-Polyakov monopole~\cite{Polyakov,Hooft} found in Yang-Mills theory coupled to a Higgs field, and the non-Abelian monopole introduced by Yang~\cite{Yang}, which can exist in five dimensions (5D). Importantly, all these monopoles carry a quantized ``magnetic" charge, which has a topological origin. Specifically, the monopole charge can be related to a topological invariant~\cite{WuYang_PRD}, which is given by the real-space integral of the curvature (or field strength tensor) associated with the monopole's gauge field~\cite{Nakahara}. This topological invariant corresponds to the ``first Chern number" in the case of three-dimensional (3D) monopoles~\cite{WuYang_PRD}, while the ``second Chern number" appears as the relevant invariant in 5D~\cite{Yang}.

The monopoles mentioned above are all associated with \emph{vector} gauge fields, an example of which is the well-known electromagnetic gauge potential~\cite{WuYang_PRD}. However, it was suggested that \emph{tensor} gauge fields could also be defined and that these more exotic gauge structures could also give rise to monopoles~\cite{Nepomechie,Teitelboim,Orland}. In this distinct class of monopoles, the simplest representative is the so-called ``tensor monopole":~an Abelian monopole defined in a four-dimensional (4D) space and whose magnetic charge is given by the integral of the curvature associated with a tensor gauge field~\cite{Kalb-Ramond}, which represents a direct generalization of the electromagnetic potential~\cite{Henneaux}. It also plays an important role in string theory, where currents naturally couple to a tensor gauge field \cite{Banks,Mavromatos,Montero}.

\begin{figure}[htp]
	\begin{center}
		\includegraphics[scale=0.55]{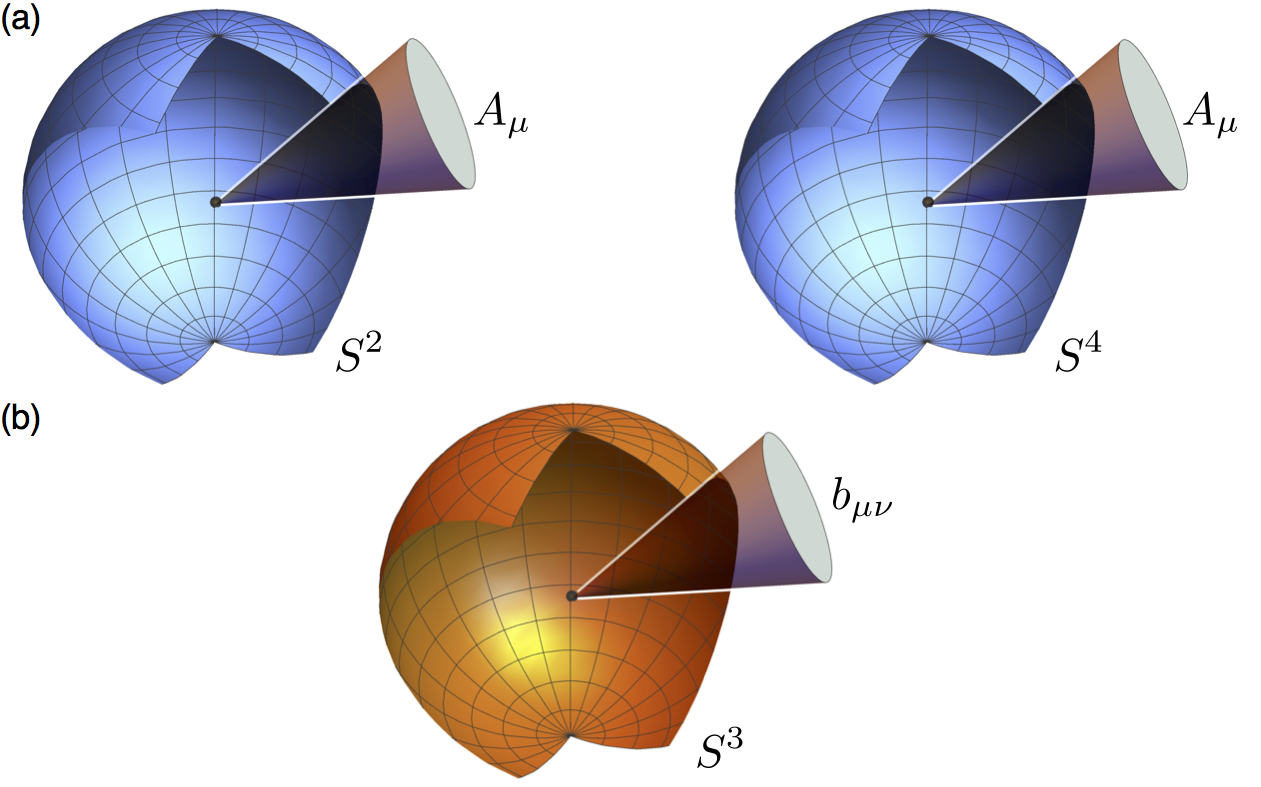}\hspace{0.5cm}
	\end{center}
	\caption{\label{Fig1} (a) Pictorial representations of a Dirac monopole in 3D (left), and of a Yang monopole in 5D (right). Both are defined as point-like objects, which are sources of a vector gauge field $A_{\mu}$. The total flux associated with $A_{\mu}$, through the surrounding spheres ($S^{2}$ and $S^{4}$, respectively) is quantized in terms of the first (resp. second) topological Chern number. (b) A tensor monopole is a point-like source of a generalized tensor gauge field $b_{\mu \nu}$, which can exist in a 4D space. The corresponding flux that comes out of the surrounding sphere $S^3$ is also quantized [Eq.~\eqref{topological_charge_generalized}].}
\end{figure}

The search for monopoles is of fundamental relevance in high-energy physics and their detection still constitutes a severe challenge. So far, none of these topological objects has been identified and there is no concrete evidence that these could be accessed in current experiments. However, monopoles also naturally appear in condensed-matter physics, where various forms of effective gauge potentials can emerge. A prominent example is the so-called Berry connection, which plays the role of a gauge potential in the parameter space of a quantum system, and which is responsible for the geometric (Berry) phase~\cite{Xiao_review,Nakahara}. Importantly,  the  monopoles associated with the Berry connection were shown to be deeply connected to the existence of topological states of matter~\cite{Fradkin,Qi_review}; see also Ref.~\cite{Castelnovo} on monopoles in spin ice. For instance, the topological invariant associated with the quantum Hall effects~\cite{Xiao_review,Qi_review} and the recently discovered Weyl semimetals~\cite{Turner,Weyl_review}, the so-called Chern number, can be simply attributed to fictitious Dirac monopoles defined in momentum space. Moreover, artificial Dirac monopoles have been implemented in ultracold atoms~\cite{Hall,Spielman}, where the high control over the physical parameters allow for tunable synthetic gauge potentials~\cite{Dalibard_review,Goldman_review}. 


The goal of this work is twice. First, we demonstrate how monopoles associated with the Berry connection of a  quantum system can be directly extracted from the quantum metric tensor (or Fubini-Study/Bures metric)~\cite{Provost}, following an approach initially developed in high-energy physics for treating real-space monopoles~\cite{Nepomechie,Freund}. This result is particularly useful for systems where the Berry curvature cannot be measured or calculated explicitly, as we further discuss below. We remind the reader that the Bures metric is related to several physical observables  \cite{Resta,Roy,Gritsev,Ozawa,Palumbo1} and that it could be directly measured in cold atoms~\cite{Ozawa-Goldman}.

Secondly, we introduce a minimal three-level model realizing a tensor monopole in a 4D parameter space. This model, which is inspired by the recent proposal~\cite{Zhang} for realizing spin-1 monopoles, could be realized in ultracold gases by coupling three internal states of an atom. In particular, it generalizes in a non-trivial way the recent experimental setting that realized non-Abelian Yang monopoles in cold atoms~\cite{Spielman}:~As stated above, the tensor monopole resulting from our model is defined in a 4D parameter space and its topological charge is related to the existence of a generalized Berry curvature associated with a tensor gauge field. In this work, we discuss how this exotic topological object could be extracted from quantum-metric measurements~\cite{Ozawa-Goldman}.
Moreover, our model exhibits an intriguing topological state that gives rise to a non-trivial generalization of Weyl semimetals in 4D, similar to that recently proposed in Ref.~\cite{Mathai}. This work represents a first step towards the analysis and implementation of novel classes of monopoles and higher-dimensional topological states of matter in quantum engineered systems~\cite{Aidelsburger_review}.

\emph{Monopoles and the quantum metric: } We start by recalling notions related to the Dirac monopole in the context of electromagnetism. Defining the electromagnetic potential $a_{\mu}$, the charge of a monopole located in $\mathbb{R}^3$ is obtained by integrating the Faraday tensor $\mathcal F_{\mu\nu}\!=\!\partial_{\mu}a_{\nu}-\partial_{\nu}a_{\mu}$ over a sphere $S^{2}$ that surrounds it; see Fig.~1 (a).  This non-vanishing flux identifies the charge, $(1/2\pi)\int_{S^2} \mathcal F\!=\!\nu^1$, which is quantized in terms of the topological Chern number $\nu^1$; here we introduced the 2-form $\mathcal{F}\!=\!(1/2)\mathcal F_{\mu\nu}\text{d}x^{\mu}\wedge\text{d}x^{\nu}$ and the wedge ($\wedge$) product~\cite{Nakahara}. Since the magnetic field emanating from a monopole is purely radial, the calculation of the total flux through $S^{2}$ essentially reduces to calculating the sphere's surface. This suggests an interesting relation between the Faraday tensor associated with a monopole and the determinant of the metric tensor $G_{\mu\nu}$, which is defined on a sphere surrounding it~\cite{Nepomechie,Freund}:
\begin{eqnarray}\label{DiracH}
F_{\mu\nu}= \epsilon_{\mu\nu}\, k \sqrt{G},
\end{eqnarray}
where $k$ is a suitable normalization constant, $\epsilon_{\mu\nu}$ is the Levi-Civita symbol and $G\!=\! \det G_{\mu\nu}$. Note that the expression~\eqref{DiracH} for the Faraday tensor is still gauge invariant and antisymmetric. Remarkably, it was demonstrated that such a relation between the Faraday tensor and a sphere's metric is not specific to the Dirac monopole: it can be systematically generalized to any point-like or extended monopoles, in any spatial dimensions~\cite{Nepomechie,Freund}. In particular, this approach is particularly useful to identify tensor monopoles~\cite{Nepomechie}, as will be illustrated below.

We first describe how this approach applies to fictitious Dirac monopoles, as defined in the parameter space of a quantum system. Consider an eigenstate $\vert u (\boldsymbol{q}) \rangle$ of a quantum system defined over some 3D parameter space spanned by $\boldsymbol{q}$ (e.g.~a state in a given Bloch band with quasi-momentum $\boldsymbol{q}$). The geometric properties of this state is captured by the quantum geometric tensor, which can be split into real and imaginary parts~\cite{Provost,Roy,Gritsev}: $\chi_{\mu\nu}\!=\!g_{\mu\nu}+\,(i/2)\Omega_{\mu\nu}$, where $\Omega_{\mu\nu}\!=\!\partial_{\mu}A_{\nu}-\partial_{\nu}A_{\mu}$ is the Berry curvature,  $A_{\mu}\!=i\!\langle u \vert \partial_{\mu}u \rangle$ is the (Abelian) Berry connection, and where 
\begin{align}
g_{\mu\nu}=&\frac{1}{2} (\langle \partial_{\mu}u  \vert \partial_{\nu} u \rangle+\langle \partial_{\nu}u \vert \partial_{\mu}u \rangle \notag \\
&- \langle \partial_{\mu}u \vert u \rangle \langle u \vert \partial_{\nu}u \rangle- \langle \partial_{\nu}u \vert u \rangle \langle u \vert \partial_{\mu}u \rangle ),\label{quantum_metric}
\end{align}
is the quantum metric tensor~\cite{Provost,Roy,Gritsev} (or Bures metric); here all derivatives are taken with respect to the parameters, i.e.~$\partial_{\mu}=\partial_{q_{\mu}}$. While the Berry curvature $\Omega$ is associated with the geometric (Berry) phase~\cite{Xiao_review}, and can be viewed as a Faraday tensor in $\boldsymbol{q}$-space~\cite{Xiao_review,Aidelsburger_review}, the quantum metric measures the (infinitesimal) distance between two nearby quantum states in $\boldsymbol{q}$-space; see Refs.~\cite{Resta,Roy,Ozawa,Gritsev,Palumbo1,Raoux,Raoux2,Gao,Ozawa-Goldman} for physical manifestations of the quantum metric.

In analogy with electromagnetism, a finite Chern number $\nu^1\!=\!(1/2\pi)\int_{S^2} \Omega$ signals the presence of a fictitious monopole in $\boldsymbol{q}$-space~\cite{Weyl_review}, located inside some sphere $S^2$. Such objects appear in the context of Weyl semimetals, where the vector $\boldsymbol{q}$ represents the quasi-momentum of a lattice system~\cite{Weyl_review}. Inspired by Eq.~\eqref{DiracH}, we now propose that such monopoles can also be detected through the determinant of a metric:~the quantum metric tensor in Eq.~\eqref{quantum_metric}. To illustrate this approach, we consider a minimal model realizing a monopole in parameter space: the Weyl Hamiltonian~\cite{Weyl_review}
\begin{eqnarray}
H_{\text{3D}}= q_{x}  \sigma^{x}+ q_{y} \sigma^{y}  + q_{z} \sigma^{z} ,
\end{eqnarray}
where $\boldsymbol{q}=(q_x,q_y,q_z)$ denotes the momentum and where $\sigma^{x,y,z}$ are the Pauli matrices. The corresponding energy spectrum supports a Weyl cone,  $E\!=\!\pm\sqrt{q_{x}^{2}+q_{y}^{2}+q_{z}^{2}}$, and the Berry curvature associated with the low-energy eigenvector $\vert u_- (\boldsymbol{q}) \rangle$ reads
\begin{eqnarray}
\Omega\!=\!\Omega_{\mu\nu} \, dq_{\mu} \wedge dq_{\nu}, \,\,\,  \Omega_{\mu\nu}\!=\!\epsilon_{\mu\nu\lambda}\,\frac{q_{\lambda}}{2(q_{x}^{2}+q_{y}^{2}+q_{z}^{2})^{3/2}}.\,  \label{curvature_weyl}
\end{eqnarray}
This corresponds to the Faraday tensor associated with a fictitious Dirac monopole at $\boldsymbol{q}\!=\!0$, as can be verified by calculating the corresponding charge; see Eq.~\eqref{charge_weyl} below.  Inspired by Eq.~\eqref{DiracH} and the approach of Refs.~\cite{Nepomechie,Freund}, we now show that this topological charge can be obtained from the determinant of the quantum metric tensor. Introducing the spherical coordinates $q_x\!=\!r \sin\theta \cos \phi$, $q_y\!=\!r \sin\theta \sin \phi$, $q_z\!=\!r \cos \theta$, the components of the quantum metric tensor~\eqref{quantum_metric} associated with the eigenvector $\vert u_- (\boldsymbol{q}) \rangle$ read~\cite{Ozawa-Goldman}
\begin{equation}
g_{\theta \theta}=1/4, \quad g_{\phi \phi} = \sin^2\theta/4 , \quad g_{\theta \phi}=0.\label{metric_weyl}
\end{equation}
This corresponds to the metric of a sphere $S^2$, of fixed radius $r\!=\!1/2$, which surrounds the Weyl node (i.e.~the monopole) in $\boldsymbol{q}$-space. Hence, the topological charge $Q$ of the Dirac monopole is encoded in the quantum metric through the relation 
\begin{eqnarray}
Q=\frac{1}{2\pi}\int_{S^{2}} \Omega=\frac{1}{2\pi}\iint (2\sqrt{g}) \, d\theta d\phi = 1,\label{charge_weyl}
\end{eqnarray}
where $g\!=\!\det g_{a b}$ is the determinant of the $2\times2$ metric tensor in Eq.~\eqref{metric_weyl}, with $a,b\!=\!\{\theta, \phi\}$. Going back to Cartesian coordinates, we find a direct relation between the components of the Berry curvature in Eq.~\eqref{curvature_weyl} and the quantum metric [Eq.~\eqref{quantum_metric}] associated with $\vert u_- (\boldsymbol{q}) \rangle$,
\begin{eqnarray}
\Omega_{\mu\nu}=\epsilon_{\mu\nu}\, (2 \sqrt{\bar g}),\label{weyl_relation}
\end{eqnarray}
where $\bar g\!=\!\det g_{\bar \mu \bar \nu}$ is the determinant of the $2\times2$ quantum metric tensor defined in the proper 2D subspace (e.g.~$\bar \mu,\bar \nu\!=\!\{q_x,q_y\}$ for the calculation of $\Omega_{xy}$). Equation~\eqref{weyl_relation} is the direct analogue of Eq.~\eqref{DiracH}, when applied to the Berry curvature, and it indicates how quantum-metric measurements~\cite{Gritsev,Ozawa-Goldman} could directly reveal Dirac monopoles in $\boldsymbol{q}$-space. It is the scope of the next paragraph to apply such an approach to a 4D Weyl-like Hamiltonian, supporting a tensor monopole~\cite{Nepomechie}.


\begin{figure}[htp]
	\begin{center}
		\includegraphics[scale=0.33]{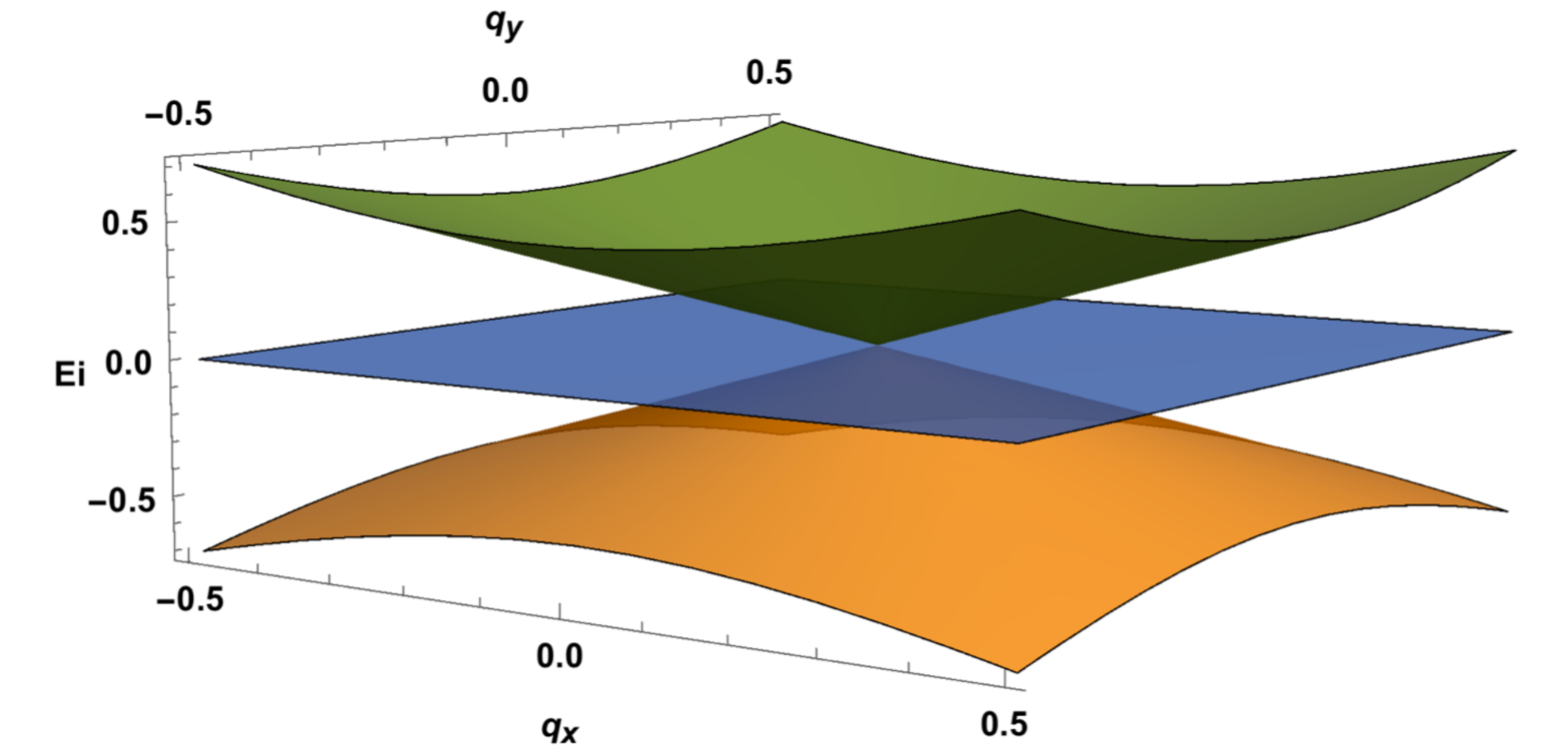}
	\end{center}
	\caption{\label{Fig2} Energy dispersion of the three-level Hamiltonian $H_{\text{4D}}$ in Eq.~\eqref{4D}, at $q_{z}\!=\!q_{w}\!=\!0$. One eigenvalue of the Hamiltonian is always zero and there appears a degenerate triple point at $\boldsymbol{q}_{\text{D}}\!=\!(0,0,0,0)$, where the tensor monopole lies. This three-band configuration, with a single degenerate triple point, offers a minimal setting realizing an Abelian tensor monopole~\cite{note_minimal}. This spectrum is also reminiscent of that found in 3D spin-1 Weyl semimetals~\cite{Burrello} and spin-1 monopoles~\cite{Zhang}. }
\end{figure}

\emph{Tensor monopoles in 4D parameter space: }
As recalled in the previous paragraph, Weyl nodes are sources of Dirac monopoles in 3D parameter space~\cite{Weyl_review}. This phenomenon can be generalized to Dirac-like nodes exhibiting a triple-point crossing, which lead to so-called spin-1 monopoles~\cite{Burrello,Zhang}; see Fig.~2. Regarding higher dimensions, Dirac nodes appearing in 5D parameter space were shown to be associated with  non-Abelian Yang monopoles~\cite{Spielman,Mathur,Avron}; see also \cite{Zhang2} on extended monopoles in 5D. Importantly, all these monopoles defined in 3D and 5D spaces are described by conventional (vector) Berry connections~\cite{Xiao_review}, i.e.~vector gauge fields. In this paragraph, we explore monopoles defined in 4D spaces, which are captured by \emph{tensor Berry connections}~\cite{note}, i.e.~tensor gauge fields~\cite{Nepomechie}.

Tensor gauge fields have been mainly analyzed in the context of high-energy physics~\cite{Nepomechie,Teitelboim,Henneaux,Kalb-Ramond,Banks,Mavromatos,Montero}, with some applications in topological states of matter~\cite{Hansson,Moore,Simon,McGreevy,Gaiotto,Hasebe,Palumbo2,Ryu,Fradkin2,Cappelli}.
The idea consists in introducing an (Abelian) antisymmetric tensor field $b_{\mu\nu}$, which naturally generalizes the usual electromagnetic potential $a_{\mu}$ \cite{Henneaux}. It transforms under a $U(1)$ gauge transformation as follows
\begin{eqnarray}
b_{\mu\nu}\rightarrow b_{\mu\nu}+\partial_{\mu}\xi_{\nu}-\partial_{\nu}\xi_{\mu},\label{gauge_transf_tensor}
\end{eqnarray}
where $\xi_{\mu}$ is a vector that contains the redundant gauge degree of  freedom of the field $b_{\mu\nu}$. The corresponding (3-form) curvature tensor has components given by
\begin{eqnarray}
C_{\mu\nu\lambda}=\partial_{\mu}b_{\nu\lambda}+\partial_{\nu}b_{\lambda\mu}+\partial_{\lambda}b_{\mu\nu}.\label{gen_curvature}
\end{eqnarray}
It is gauge invariant and antisymmetric, in analogy with the Faraday tensor.
As shown in Refs.~\cite{Nepomechie,Teitelboim}, $b_{\mu\nu}$ gives rise to a novel type of monopole in 4D space, which is Abelian and point-like. In this sense, this tensor monopole generalizes the 3D Dirac monopole to 4D. Specifically, if one surrounds the tensor monopole with a three-dimensional sphere $S^{3}$, see Fig.~1(b), one can derive a topological  charge associated with the (3-form) curvature $C_{\mu\nu\lambda}$,
\begin{eqnarray}
Q_{T}=\frac{1}{2\pi^2}\int_{S^{3}}dx^{\mu}\wedge dx^{\nu}\wedge dx^{\lambda}\,C_{\mu\nu\lambda} .\label{topological_charge_generalized}
\end{eqnarray}
This is a topological invariant known as the Dixmier-Douady invariant~\cite{Mathai,Murray1,Murray2}, a generalization of the better-known Chern number~\cite{Nakahara}; this invariant, which is associated with the third homotopy group $\pi_3(S^3)\!=\!\mathbb{Z}$, characterizes a U(1) ``bundle gerbe" ~\cite{math_notes,math_notes2}.
Importantly, the result in Eq.~\eqref{DiracH} can be directly generalized to tensor monopoles~\cite{Nepomechie}, which indicates that the topological charge in Eq.~\eqref{topological_charge_generalized} can be calculated through the determinant of the metric tensor defined on $S^{3}$.

The goal of this paragraph is to show how the parameter-space analogue of the curvature $C_{\mu\nu\lambda}$ [Eq.~\eqref{gen_curvature}], as well as its topological charge [Eq.~\eqref{topological_charge_generalized}], can be determined through the quantum metric tensor of a quantum system. To do so, we consider a minimal Weyl-type Hamiltonian defined in a 4D parameter space~\cite{note_minimal}, spanned by $\boldsymbol{q}\!=\!(q_{x},q_{y},q_{z},q_{w})$,
\begin{align}\label{4D}
H_{\text{4D}}&=q_{x}\lambda_{1}+q_{y}\lambda_{2}+q_{z}\lambda_{6}+q_{w}\lambda_{7}^{*}, \\
&= \left({\begin{array}{ccc}
	0 & q_{x}- i q_{y} & 0 \\
	q_{x}+ i q_{y} & 0 & q_{z}+ i q_{w} \\
	0 & q_{z}- i q_{w} & 0 \\
	\end{array} } \right), \nonumber
\end{align}
where the $\lambda$ matrices are $3\times3$ Gell-Mann matrices~\cite{Gell-Mann}.

The corresponding spectrum is given by
\begin{eqnarray}
E_{0}=0, \hspace{0.5cm} E_{\pm}=\pm\sqrt{q_{x}^{2}+q_{y}^{2}+q_{z}^{2}+ q_{w}^{2}},
\end{eqnarray}
where we recognize the presence of a triple-degenerate Dirac-like point at $\boldsymbol{q}_{\text{D}}=(0,0,0,0)$;  see Fig.~2. We now demonstrate that this 4D Dirac-like node is a source of a tensor monopole, based on the approach developed in the previous paragraph: Inspired by Ref.~\cite{Nepomechie} and Eqs.~\eqref{weyl_relation},\eqref{gen_curvature}, we write the generalized Berry curvature tensor as
\begin{eqnarray} \label{H}
W_{\mu \nu \lambda}=\epsilon_{\mu\nu\lambda} \, (4 \sqrt{\bar{g}}),
\end{eqnarray}
where $\bar g\!=\!\det g_{\bar \mu \bar \nu}$ is the determinant of the $3\times3$ metric tensor defined in the proper 3D subspace (e.g.~$\bar \mu,\bar \nu\!=\!\{q_x,q_y,q_z\}$ for the calculation of $W_{xyz}$). This quantum metric can be directly evaluated using Eq.~\eqref{quantum_metric} and the expression for the low-energy eigenstate of $H_{\text{4D}}$ in Eq.~\eqref{4D}, 
\begin{eqnarray}
\vert u_{-} \rangle=\frac{1}{\sqrt{2}}(v_{1},-1,v_{2})^{\top}, \, v_{1}=\frac{q_{x}-i q_{y}}{|E|}, \, v_{2}=\frac{q_{z}-i q_{w}}{|E|}.\notag
\end{eqnarray}
Similarly to the case of the 3D Weyl Hamiltonian [Eq.~\eqref{metric_weyl}], we find that the corresponding quantum metric identifies a three-sphere $S^{3}$, with fixed radius, surrounding the Dirac-like node. Furthermore, when combined with the ansatz in Eq.~\eqref{H}, we obtain
\begin{eqnarray}
W_{\mu\nu\lambda}=\epsilon_{\mu\nu\lambda\gamma}\,\frac{q_{\gamma}}{(q_{x}^{2}+q_{y}^{2}+q_{z}^{2}+q_{w}^{2})^{2}},\label{eq:3curvature}
\end{eqnarray}
which indeed coincides with the curvature of a tensor monopole in 4D~\cite{Nepomechie}.
Introducing the hyper-spherical coordinates,
$(r,\theta_{1},\theta_{2},\varphi)$, see \cite{HyperS}, 
we now explicitly calculate the topological charge $Q_{T}$ associated with $W_{\mu\nu\lambda}$,
\begin{align}\label{QT}
Q_{T}&=\frac{1}{2\pi^2}\int_{S^{3}}dq^{\mu}\wedge dq^{\nu}\wedge dq^{\lambda}\,W_{\mu\nu\lambda} \\ 
&=\frac{1}{2\pi^{2}}\int_{0}^{\pi}d\theta_{1}\int_{0}^{\pi}d\theta_{2}\int_{0}^{2\pi}d\varphi \sin^{2} \theta_{1} \sin \theta_{2} =1.\notag
\end{align}
We emphasize that this topological charge was directly identified through the quantum metric, that is, without relying on any expression for the tensor Berry connection [i.e.~the gauge field $b_{\mu \nu}$ in Fig.~1 (b)]; an alternative approach, based on the explicit evaluation of the Berry connection, is provided in Refs.~\cite{sup_mat,preparation}. One verifies that the charge $Q_{T}$ is immune to smooth deformations of the Hamiltonian that preserve the Dirac-like node, in agreement with its topological nature.

If we identify the parameters $\boldsymbol{q}$ with the quasi-momenta of a suitable 4D lattice model, Eq.~\eqref{4D} can be seen as the linearized Hamiltonian of a generalized Weyl semimetal defined in 4D, similar to that proposed in Ref.~\cite{Mathai}. Moreover, by taking a slice of our 4D model, at some fixed $q_{w}\!=\!{\rm const.}$, we obtain a three-dimensional gapped phase, which describes a topological insulator in the chiral class AIII; indeed, the Hamiltonian $H_{\text{4D}}$ preserves chiral symmetry according to
\begin{eqnarray}
UH_{\text{4D}}(\textbf{q})U^{-1}= -H_{\text{4D}}(\textbf{q}), \hspace{0.2cm}
U= \left({\begin{array}{ccc}
	1 & 0 & 0 \\
	0 & 0 & 1 \\
	0 & -1 & 0 \\
	\end{array} } \right).\notag
\end{eqnarray}
Such AIII topological phases were recently explored in a 3D model~\cite{Neupert}. In principle, 4D lattice models leading to the Hamiltonian in Eq.~\eqref{4D} could be realized in optical lattices, e.g.~using synthetic dimensions~\cite{Goldman2,Goldman3}. However, in the next paragraph, we will present a simpler implementation of this Hamiltonian through a three-level system of ultracold atoms.

\emph{Ultracold-atoms implementation: } In this paragraph, we present a possible physical realization of a tensor monopole, through the manipulation of three atomic levels. A natural choice would be three sublevels within the hyperfine ground-states of $^{87}$Rb atoms, coupled by two (RF or microwave) driving fields~\cite{Spielman,Zhang}. The corresponding three-level Hamiltonian can be written in the general form
\begin{eqnarray}\label{BEC}
H_{\text{exp}}= \left({\begin{array}{ccc}
	-\delta_{12} & \omega_{12}e^{-i \phi_{12}} & 0 \\
	\omega_{12}e^{i \phi_{12}} & 0 & \omega_{23}e^{i \phi_{23}} \\
	0 & \omega_{23}e^{-i \phi_{23}} & \delta_{23} \\
	\end{array} } \right),
\end{eqnarray}
where $\omega_{ij}$ and $\phi_{ij}$ are the Rabi amplitudes and phases of  the coupling fields~\cite{Dalibard_review,Goldman_review}, respectively, and where $\delta_{ij}$ capture the detuning from resonances. In contrast with previous proposals~\cite{Zhang}, we now neglect these detuning effects ($\delta_{ij}\!=\!0$), which we assume to be small compared to the Rabi amplitudes. In this regime, the Hamiltonian $H_{\text{exp}}$ depends on four independent parameters only, and it can therefore be mapped unto the Hamiltonian in Eq.~\eqref{4D}, through the following identifications
\begin{eqnarray}
\omega_{12}e^{i \phi_{12}}=q_{x}+ i q_{y}, \hspace{0.5cm} \omega_{23}e^{i \phi_{23}}=q_{z}+ i q_{w}.
\end{eqnarray}
Hence, this minimal platform supports a (fictitious) tensor monopole in the parameter space spanned by $\{\omega_{12}, \phi_{12}, \omega_{23}, \phi_{23}\}$, and it is characterized by a non-zero topological charge $Q_{T}$; see Eq.~\eqref{QT}. As shown in the previous paragraph, this topological invariant can be directly obtained from the quantum metric associated with the Hamiltonian's eigenstates. Following the protocol of Ref.~\cite{Ozawa-Goldman}, the components of the quantum metric could be individually obtained by initially preparing the system in the low-energy eigenstate of the Hamiltonian~\eqref{BEC}, and then monitoring the excitation rate upon modulating the system parameters in time~\cite{sup_mat}; see Ref.~\cite{Gritsev} for other possible probes of the metric. This would represent a direct method to reveal, for the first time, the existence of tensor monopoles through the identification of their monopole charge via quantum-metric measurements.

%

 
{\bf Acknowledgments: }
We are grateful to M. Di Liberto for a careful reading of a preliminary version of the manuscript 
and to M. Aidelsburger, B. Mera and T. Ozawa for helpful discussions. This work was supported by the FRS-FNRS (Belgium) and the ERC Starting Grant TopoCold. \\


\newpage

\begin{center}

\Large{Supplemental Material}

\end{center}

\subsection{The tensor Berry connection of a 4D monopole}

In the main text, we investigated the geometric and topological structures associated with a 4D tensor monopole, which was defined in the 4D parameter space of a quantum system described by the Weyl-type Hamiltonian in Eq.~\eqref{4D}. The approach developed in the main text was based on Eq.~\eqref{H}, namely, an expression that relates the 3-form curvature $W_{\mu \nu \lambda}$ characterizing the tensor monopole to the quantum metric $g_{\mu \nu}$ associated with the eigenstates of the Hamiltonian. While this approach provides an explicit expression for the curvature [Eq.~\eqref{eq:3curvature}], and therefore an estimation of the related topological charge, it disregards an important aspect of the geometric structure, namely, the underlying 2-form connection $b_{\mu\nu}$: the fundamental tensor gauge field that is at the origin of the curvature. 

In this Appendix, we fill this gap by deriving an explicit expression for the underlying tensor connection $b_{\mu\nu}$, namely, the antisymmetric tensor field whose external derivative generates the curvature
\begin{eqnarray}
W_{\mu\nu\lambda}=\partial_{\mu}b_{\nu\lambda}+\partial_{\nu}b_{\lambda\mu}+\partial_{\lambda}b_{\mu\nu} ,\label{gen_curvature_appendix}
\end{eqnarray}
where $W_{\mu\nu\lambda}$ is the 3-form curvature obtained in the main text [Eq.~\eqref{eq:3curvature}]. Importantly, this second approach allows us to characterize the topology of the 4D monopole without relying on the calculation of the quantum metric nor the use of Eq.~\eqref{H}. In fact, the following discussion constitutes a short introduction to the concept of ``tensor Berry connection", which will be discussed in full detail in Ref.~\cite{preparation}. 

Let us first remind that an Abelian tensor gauge field satisfies the gauge transformation, or local shift symmetry~\cite{Dvali},
\begin{eqnarray}\label{BB-field}
b_{\mu\nu}\rightarrow b_{\mu\nu}+\Lambda_{\mu\nu},
\end{eqnarray}
where $\Lambda_{\mu\nu}$ is an arbitrary 2-form field. In specific cases, $\Lambda_{\mu\nu}$ can itself be expressed as an exterior derivative $\Lambda_{\mu\nu}\!=\!\partial_{\mu}\xi_{\nu}\!-\!\partial_{\nu}\xi_{\mu}$, where $\xi_{\mu}$ is a 1-form; see Eq.~\eqref{gauge_transf_tensor} in the main text.

Inspired by Ref.~\cite{Pacheva}, we start by introducing a set of three scalar fields $\phi_{1,2,3} (\bs q)$, which depend on the parameters $\bs q\!=\!(q_{x},q_{y}, q_{z}, q_{w})$ of the system [Eq.~\eqref{4D} in the main text], and from which we build an associated tensor gauge field as
\begin{eqnarray}\label{tensorB}
b_{\mu\nu}=\frac{i}{3}\,\epsilon^{jkl} \phi_{j}\partial_{\mu}\phi_{k}\partial_{\nu}\phi_{l}, \qquad j,k,l=\{1,2,3\},
\end{eqnarray}
where $\mu,\nu\!=\!\{q_{x},q_{y}, q_{z}, q_{w}\}$ refer to the parameters of the system. We note that the expression in Eq.~\eqref{tensorB} indeed describes a completely antisymmetric tensor $b_{\mu\nu}\!=\!-b_{\nu\mu}$. Next, we impose that the three scalars should satisfy the following local gauge transformations
\begin{eqnarray}\label{gauge7}
\phi_{1}\rightarrow \phi_{1}+\alpha(\bs q),\hspace{0.1cm}\phi_{2}\rightarrow e^{-i \alpha(\bs q)} \phi_{2}, \hspace{0.1cm}
\phi_{3}\rightarrow e^{i \alpha(\bs q)} \phi_{3},
\end{eqnarray}
where $\alpha(\bs q)$ is a function of the system parameters. Under those assumptions, one verifies that the 2-form $b_{\mu\nu}$ defined in Eq.~\eqref{tensorB} indeed transforms as a tensor gauge field [Eq.~\eqref{BB-field}] under the gauge transformations [Eq.~\eqref{gauge7}]; here the resulting shift $\Lambda_{\mu\nu}$ is a function of the fields $\phi_{i}$ and $\alpha$, namely, it contains the gauge redundancy of the scalars.

Let us now connect these notions to the eigenstates of our 4D Weyl-type Hamiltonian (see main text)
\begin{eqnarray}
\vert u_{-} \rangle=\frac{1}{\sqrt{2}}(v_{1},-1,v_{2})^{\top}, \, v_{1}=\frac{q_{x}-i q_{y}}{|E|}, \, v_{2}=\frac{q_{z}-i q_{w}}{|E|}.\notag
\end{eqnarray}
Noting that these eigenstates transform as $|u_{-}\rangle\!\rightarrow\!e^{i \alpha(\bs q)}|u_{-}\rangle$ under a local gauge transformation, one identifies three scalar fields
\begin{eqnarray}\label{gauge8}
\phi_{1}\equiv -i\log v_{2},\hspace{0.3cm}\phi_{2}\equiv v_{1}^{*}, \hspace{0.3cm}
\phi_{3}\equiv v_{1},
\end{eqnarray}
which satisfy the generalized transformations defined in Eq.~\eqref{gauge7}. When combined with Eq.~\eqref{tensorB}, the definitions in Eq.~\eqref{gauge8} eventually allows us to obtain an explicit expression for the tensor connection $b_{\mu\nu}(\bs q)$ in terms of the eigenstates components $v_{1,2}$. After a tedious calculation, one verifies that this ``tensor Berry connection" indeed generates the curvature
\begin{align}
W_{\mu\nu\lambda}&=\partial_{\mu}b_{\nu\lambda}+\partial_{\nu}b_{\lambda\mu}+\partial_{\lambda}b_{\mu\nu} \nonumber \\
&=\epsilon_{\mu\nu\lambda\gamma}\,\frac{q_{\gamma}}{(q_{x}^{2}+q_{y}^{2}+q_{z}^{2}+q_{w}^{2})^{2}},
\end{align}
as previously derived in the main text through the quantum metric [see Eq.~\eqref{eq:3curvature}].

\subsection{Measuring a tensor monopole in a three-level system}

This second part of the Appendix aims to provide more details regarding the realization and detection of a tensor monopole using a three-level system. We start by describing the general setting and then discuss how the quantum-metric measurement of Ref.~\cite{Ozawa-Goldman} applies to this situation. 

Let us start by considering a three-level system, involving the states $\{\vert 1 \rangle, \vert 2 \rangle , \vert 3 \rangle  \}$ and described by the Hamiltonian
\be
\hat H_0=\varepsilon_3 \vert 3 \rangle \langle 3 \vert - \varepsilon_1 \vert 1 \rangle \langle 1 \vert .
\ee
Here, we explicitly set the energy of the second level to zero, $\varepsilon_2\!=\!0$, and we assume that $\varepsilon_1,\varepsilon_3\!>\!0$ in the following. We then introduce coupling between these levels, which we take in the form of two time-dependent terms:
\begin{align}
&\hat V_{12}(t)= \hat V e^{i (\theta_{12} t - \phi_{12})} + \hat V^{\dagger} e^{-i (\theta_{12} t - \phi_{12})},\\
&\hat V_{23}(t)= \hat W e^{i (\theta_{23} t + \phi_{23})} + \hat W^{\dagger} e^{-i (\theta_{23} t + \phi_{23})},
\end{align}
where $\theta_{12,23}$ and $\phi_{12,23}$ denote the corresponding frequencies and phases. For the sake of simplicity, we assume that all coupling matrix elements associated with the operators $\hat V$ and $\hat W$  are zero, except
\begin{align}
\langle 1 \vert \hat V \vert 2 \rangle = \langle 2 \vert \hat V \vert 1 \rangle \equiv \Omega_{12}^*, \qquad \langle 2 \vert \hat W \vert 3 \rangle =\langle 3 \vert \hat W \vert 2 \rangle \equiv \Omega_{23},\notag
\end{align}
and we note that the Rabi amplitudes $\Omega_{12}$ and $\Omega_{23}$ are complex numbers (these quantities should not be confused with the Berry curvature $\Omega_{\mu \nu}$ discussed in the main text). In a rotating frame generated by the unitary operator
\be
\hat R (t)= \exp \left ( i \left [ \theta_{23} t \vert 3 \rangle \langle 3 \vert - \theta_{12} t \vert 1 \rangle \langle 1 \vert  \right ] \right ),
\ee
the total Hamiltonian $\hat H (t)\!=\!\hat H_0\!+\!\hat V_{12}(t)\!+\!\hat V_{23}(t)$ eventually takes the form of an effective Hamiltonian
\begin{align}
\hat H&=\delta_{23} \vert 3 \rangle \langle 3 \vert - \delta_{12} \vert 1 \rangle \langle 1 \vert \\
&+ \left ( \Omega_{12}^* \vert 1 \rangle \langle 2 \vert ^{-i \phi_{12}}+ \text{h.c.}\right )+ \left ( \Omega_{23} \vert 2 \rangle \langle 3 \vert ^{i \phi_{23}}+ \text{h.c.}\right ),\notag
\end{align}
upon applying the standard rotating-wave-application (i.e.~neglecting all fast-oscillating terms); here we introduced the detunings $\delta_{12}\!=\!\varepsilon_1\!-\!\theta_{12}$ and $\delta_{23}\!=\!\varepsilon_3\!-\!\theta_{23}$. Neglecting these detuning effects ($\delta_{12}\!=\!\delta_{12}\!=\!0$), one obtains the Hamiltonian matrix
\begin{eqnarray}\label{BEC}
\hat H_{\text{exp}}= \left({\begin{array}{ccc}
	0 & \Omega_{12}^*e^{-i \phi_{12}} & 0 \\
	\Omega_{12}e^{i \phi_{12}} & 0 & \Omega_{23}e^{i \phi_{23}} \\
	0 & \Omega_{23}^*e^{-i \phi_{23}} & 0 \\
	\end{array} } \right) .
\end{eqnarray}
Considering the case where the Rabi amplitudes are real ($\Omega_{12}^*\!=\!\Omega_{12}\!\equiv\!\omega_{12}$, $\Omega_{23}^*\!=\!\Omega_{23}\!\equiv\!\omega_{23}$), the corresponding Hamiltonian 
\begin{eqnarray}\label{BECtwo}
\hat H_{\text{exp}}= \left({\begin{array}{ccc}
	0 & \omega_{12}e^{-i \phi_{12}} & 0 \\
	\omega_{12}e^{i \phi_{12}} & 0 & \omega_{23}e^{i \phi_{23}} \\
	0 & \omega_{23}e^{-i \phi_{23}} & 0 \\
	\end{array} } \right) , \label{ham_exp}
\end{eqnarray}
can be directly mapped to the 4D Weyl-type Hamiltonian [Eq.~(11) in the main text]
\begin{align}
\hat H_{\text{4D}}= \left({\begin{array}{ccc}
	0 & q_{x}- i q_{y} & 0 \\
	q_{x}+ i q_{y} & 0 & q_{z}+ i q_{w} \\
	0 & q_{z}- i q_{w} & 0 \\
	\end{array} } \right), \label{4D_ham}
\end{align}
through the simple identifications
\begin{eqnarray}
\omega_{12}e^{i \phi_{12}}=q_{x}+ i q_{y}, \hspace{0.5cm} \omega_{23}e^{i \phi_{23}}=q_{z}+ i q_{w}.
\end{eqnarray}
We note that such a configuration [Eq.~\eqref{ham_exp}] can be simply achieved by coupling three hyperfine ground states of $^{87}$Rb atoms with two RF (or microwave) driving fields; see Ref.~\cite{Spielman} for an experimental realization and Ref.~\cite{Zhang} for a related proposal. As discussed in the main text, this simple setting would realize a tensor monopole in an ultracold atom experiment.\\

The tensor monopole associated with the 4D Weyl-type Hamiltonian in Eq.~\eqref{4D_ham} can be directly extracted from the quantum metric tensor. Recently, it was demonstrated that this geometric object could be directly measured in ultracold atoms, using a universal scheme based on excitation-rate measurements~\cite{Ozawa-Goldman}; we note that this method can be applied to any multi-level (or multi-band) model in any dimensions (in contrast with other approaches based on state tomography~\cite{Flaschner,superconducting}). In short, the method of Ref.~\cite{Ozawa-Goldman} consists in preparing the system in the lowest-energy eigenstate of the Hamiltonian of interest, in our case $\hat H_{\text{4D}} (\bs q^0)$, for some value of the parameters $\bs q^0$; then by modulating a parameter in time, for instance
\be
q_{\mu} (t)=q_{\mu}^0 + (2 \mathcal E / \omega) \cos (\omega t),\label{mod_one}
\ee
one obtains the diagonal component $g_{\mu \mu}(\bs q^0)$ of the quantum metric tensor by measuring the frequency-integrated excitation rate~\cite{Ozawa-Goldman}
\be
\Gamma^{\text{int}}=\int_0^{\infty} \text{d}\omega \, \Gamma(\omega) = (2 \pi \mathcal E^2) g_{\mu \mu}(\bs q^0),
\ee
where $\Gamma(\omega)$ denotes the excitation rate upon driving the system with the frequency $\omega$, and where $\mathcal E$ is the strength of the drive in Eq.~\eqref{mod_one}. Similarly, the off-diagonal components of the quantum metric tensor, $g_{\mu \nu}(\bs q^0)$, can be obtained by modulating the two corresponding parameters in time, namely,
\begin{align}
q_{\mu} (t)=q_{\mu}^0 + (2 \mathcal E / \omega) \cos (\omega t), \notag\\
q_{\nu}^{\pm} (t)=q_{\nu}^0 \pm (2 \mathcal E / \omega) \cos (\omega t)\label{mod_two},
\end{align}
and measuring the differential integrated rate~\cite{Ozawa-Goldman}
\be
\Delta \Gamma^{\text{int}}=\int_0^{\infty} \text{d}\omega \, \left [\Gamma^+(\omega)-\Gamma^-(\omega) \right ]= (8 \pi \mathcal E^2) g_{\mu \nu}(\bs q^0),\notag
\ee
where $\Gamma^{\pm}$ refer to the excitation rates resulting from the drives $q_{\nu}^{\pm} (t)$ in Eq.~\eqref{mod_two}.

Such a protocol could be applied to the general three-level Hamiltonian in Eq.~\eqref{BEC}. Indeed, let us first notice that the latter can be directly mapped to the 4D Weyl-type Hamiltonian in Eq.~\eqref{4D_ham} through the identifications
\begin{align}
&q_x=\mathcal{R}\left [ \Omega_{12}e^{i \phi_{12}} \right ] , q_y=\mathcal{I}\left [ \Omega_{12}e^{i \phi_{12}} \right ], \notag \\
&q_z=\mathcal{R}\left [ \Omega_{23}e^{i \phi_{23}} \right ] , q_w=\mathcal{I}\left [ \Omega_{23}e^{i \phi_{23}} \right ],
\end{align}
where $\mathcal{R}/\mathcal{I}$ refer to the real and imaginary parts, and we remind that the Rabi amplitudes $\Omega_{12}$ and $\Omega_{23}$ are complex numbers in general. Then, as explained above, the protocol of Ref.~\cite{Ozawa-Goldman} requires driving a set of parameters $\{ q_{\mu}(t) , q_{\nu}(t) \}$ periodically in time, which can be realized in this setting by modulating the real and imaginary parts of the Rabi amplitudes $\Omega_{12}$ and $\Omega_{23}$ in a proper manner; see Eqs.~\eqref{mod_one} and~\eqref{mod_two}. Measuring the resulting excitation rates $\Gamma (\omega)$, for a proper range of drive frequencies $\omega$, would then provide the different components of the quantum metric tensor, and hence, reveal the tensor monopole associated with this 4D model (see main text). 

We note that the scheme above requires individual access to the real and imaginary parts of the Rabi amplitudes; while this could be achieved by coupling three internal states of an atom with a well-designed laser configuration, such a scheme could be facilitated by exploiting laser-induced tunneling~\cite{Goldman_review} in a three-site optical lattice (where each site would be associated with a fictitious level).


\begin{thebibliography}{100}
\bibitem{note}
The concept of a tensor Berry connection was also suggested in Ref.~\cite{Mathai}, but without providing any explicit expression for the corresponding curvature.
\bibitem{note_minimal} Tensor monopoles can exist in parameter spaces of dimension $D\!\ge\!4$; see Ref.~\cite{Nepomechie}. A three-band model defined in 4D, which exhibits a single degenerate triple point, constitutes a minimal setting for the realization of an Abelian tensor monopole.
\bibitem{HyperS}
The hyper-spherical coordinates are defined as 
$q_{x}\!=\!r\cos \theta_{1}$, $q_{y}\!=\!r\sin \theta_{1} \cos \theta_{2}$, $q_{z}\!=\!r \sin \theta_{1} \sin \theta_{2} \cos \varphi$, $q_{w}\!=\! r \sin \theta_{1} \sin \theta_{2}\sin \varphi$.
\end{thebibliography}
\end{document}